# Hydrogen plasma inhibits ion beam restructuring of materials

John A. Scott,[*,†] James Bishop,[‡] Garrett Budnik,[¶] and Milos Toth[*,‡,§]

†*Institute for Photonics and Optical Sciences (IPOS),School of Physics, The University of Sydney, Camperdown, NSW, 2006, Australia*

‡*School of Mathematical and Physical Sciences, University of Technology Sydney, Ultimo, New South Wales 2007, Australia*

¶*Advanced Technology, Thermo Fisher Scientific, NE Dawson Creek Dr., Hillsboro, 97124, OR, USA*

§*ARC Centre of Excellence for Transformative Meta-Optical Systems, University of Technology Sydney, Ultimo, New South Wales 2007, Australia*

E-mail: John.Scott1@sydney.edu.au; Milos.Toth@uts.edu.au

**Abstract**

Focused ion beam (FIB) techniques are employed widely for nanofabrication, and processing of materials and devices. However, ion irradiation often gives rise to severe damage due to atomic displacements that cause defect formation, migration and clustering within the ion-solid interaction volume. The resulting restructuring degrades the functionality of materials, and limits the utility FIB ablation and nanofabrication techniques. Here we show that such restructuring can be inhibited by performing FIB irradiation in a hydrogen plasma environment *via* chemical pathways that modify defect binding energies and transport kinetics, as well as material ablation rates. The method is minimally-invasive and has the potential to greatly expand the utility of FIB nanofabrication techniques in processing of functional materials and devices.

# Introduction

Focused ion beam (FIB) processing of materials plays a critical role in fields that include nanoelectronics, nanophotonics, and electron microscopy [1, 2, 3, 4, 5]. However, despite a high degree of technological maturity, the utility of FIB techniques is limited by unintended restructuring of materials by energetic (keV) ions. In particular, the compositional stoichiometry and topography of compound materials are often modified profoundly by ion impacts. This is exemplified by the well-studied III-V semiconductors GaP, GaAs, GaN, InN and InAs [6, 7, 8, 9, 10]. In these materials, ion bombardment causes segregation and accumulation of the group III element at the surface, resulting in modification of both the composition and topography (i.e., roughness) of the surface. Such ion beam restructuring compromises the functionality of materials and devices, and limits the applicability of FIB processing and nanofabrication techniques.

Here, we introduce a minimally-intrusive chemical method to inhibit restructuring caused by energetic ions. We show that hydrogen radicals delivered to the surface of a III-V semiconductor (Fig. 1(a)) can suppress restructuring caused by a FIB. Specifically, we irradiate GaP by 12 keV Xe$^+$ ions in high vacuum, and in the presence of hydrogen radicals generated by a remote plasma microinjector. In vacuum, we observe severe ion beam restructuring, which includes the accumulation of excess Ga and formation of Ga droplets at the surface (Fig. 1(b)). Conversely, in a hydrogen plasma environment, FIB irradiation yields smooth GaP surfaces devoid of excess Ga. This is attributed to hydrogen-mediated chemical pathways that modify defect kinetics during ion irradiation, and inhibit material restructuring caused by the ions.

The underlying mechanisms are elucidated using two sets of reference control experiments. First, hydrogen radicals generated by the plasma are replaced with H$_2$ gas or H$_2$O vapor to illustrate the chemical effects of these molecules on material restructuring. Second, hybrid dynamic experiments are performed in which FIB restructuring of GaP is initiated in vacuum to generate excess Ga at the surface. Hydrogen radicals, H$_2$ gas, or H$_2$O vapor are then introduced (during ion irradiation) to observe their effects, in real time, on the excess metallic Ga at the surface. Our results indicate that hydrogen radicals suppress defect migration and clustering during ion bombardment of GaP, and also increase the physical sputtering yield of Ga clusters. More broadly, the results indicate that hydrogen suppresses ion beam restructuring of materials by modifying the binding energies of crystallographic defects.

Our work shows that radicals generated by a hydrogen plasma can be used to prevent compositional restructuring of a compound material by energetic ions. The approach is demonstrated using GaP, and has potential for broad applicability to materials that are grown and processed in hydrogen-rich environments.

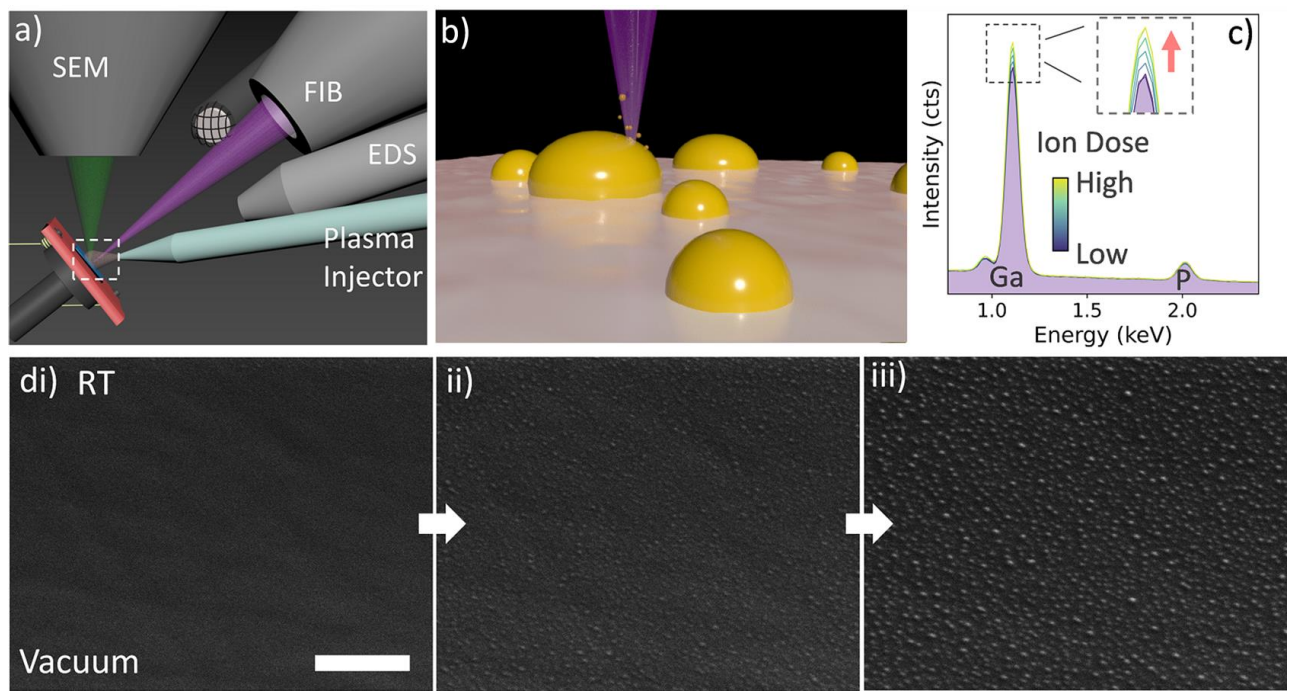

Figure 1: **Experimental setup, and Xe$^+$ ion beam irradiation of GaP in vacuum at room temperature (RT). (a)** Coincident focused ion beam (FIB), electron beam and a gas/plasma microinjector. The electron beam is used for scanning electron microscopy (SEM) and energy-dispersive x-ray spectroscopy (EDS). **(b)** Illustration of Ga droplet formation caused by ion beam irradiation of GaP. **(c)** EDS spectra taken at 10 s intervals, showing the accumulation of Ga at the surface of GaP caused by ion irradiation in vacuum. The inset is a close-up of the Ga$_{K\alpha}$ peak. **(d)** SEM images of GaP taken at FIB irradiation times of (i) 0 s, (ii) 20 s and (iii) 40 s, showing the formation of Ga nanodroplets in vacuum. The scale bar represents 500 nm. The Xe$^+$ ion beam energy and current were 12 keV and 0.8 nA. The electron beam conditions used to collect SEM images and x-ray spectra are detailed in the Methods Section.

# Results and discussion

## Ion beam irradiation in vacuum

We start by demonstrating the effects of FIB restructuring (in vacuum) on the composition and topography of ⟨100⟩-oriented GaP. The ion Xe$^+$ is employed as it is relatively inert and it is not present in pristine GaP. This allows unambiguous detection of changes in surface composition that are caused by ion beam restructuring, rather than the implantation of Ga$^+$ (the most common ion used in FIB instruments).

Irradiation of GaP by 12 keV Xe$^+$ (Figure 1(a,b)) alters the composition of the surface as is demonstrated in Figure 1(c) by energy-dispersive x-ray spectroscopy (EDS). EDS is an electron beam analysis technique that is surface-sensitive to a degree determined by the employed electron beam energy and incidence angle[11] (which were set to 3 keV and 52°, as is detailed in Methods). The spectra show that the intensity of the Ga$_{K\alpha}$ x-ray peak increases with time (i.e., Xe$^+$ ion dose), due to the build-up of excess Ga at the GaP surface during FIB irradiation. The excess Ga coalesces into droplets, as is illustrated in Figure 1(d) by a series of SEM images of a single region of GaP taken: (i) before, and (ii, iii) after cumulative irradiation by 12 keV Xe$^+$ ions. Images were acquired periodically whilst the FIB irradiation was paused at 5 s intervals. Images acquired at 0, 20 and 40 s are shown in the figure. A movie of the droplet formation process is provided in the Supplementary Information, Supplementary Video S1. The movie was produced from SEM images acquired by interrupting the FIB irradiation periodically.

The images in Figure 1(d) and Supplementary Video S1 demonstrate the formation of Ga nanodroplets caused by ion beam restructuring of GaP. Importantly, the droplet formation is not a simple consequence of preferential sputtering of P from GaP, as the sputter yields of Ga and P in GaP are 2.1 and 1.7, respectively.[12] The droplet formation in Ga-containing materials[13, 14, 15, 16, 17, 18, 19] is instead a consequence of the high cohesive energy of Ga, and proceeds through ion-induced defect migration, and associated Ga coalescence, nucleation and ripening mechanisms.[20, 9, 19] That is, ion irradiation generates defects within the ion-solid interaction volume which diffuse and interact to form clusters, including stable Ga clusters and eventually droplets with high Ga-Ga binding energies and low sputter yields. The droplets grow (under continued ion irradiation) to a critical size[20] at which the growth rate is equal to the sputter rate. We note that droplet formation is not a universal consequence of ion beam restructuring of compound materials. It is, however, a consequence of a change in surface composition (which is universal), and thus a useful model system for the present study as it can be monitored in real time by SEM imaging during ion irradiation.

Droplet formation modifies not only the surface composition, but also the topography, which we imaged *ex-situ* by atomic force microscopy (AFM). Figure 2(a) shows an AFM map of GaP that had been irradiated (in vacuum) by 12 keV Xe$^+$ ions. The droplet height can reach ∼ 10 nm, giving rise to substantial surface roughness, as is illustrated by the pink AFM line profile in Figure 2(d).

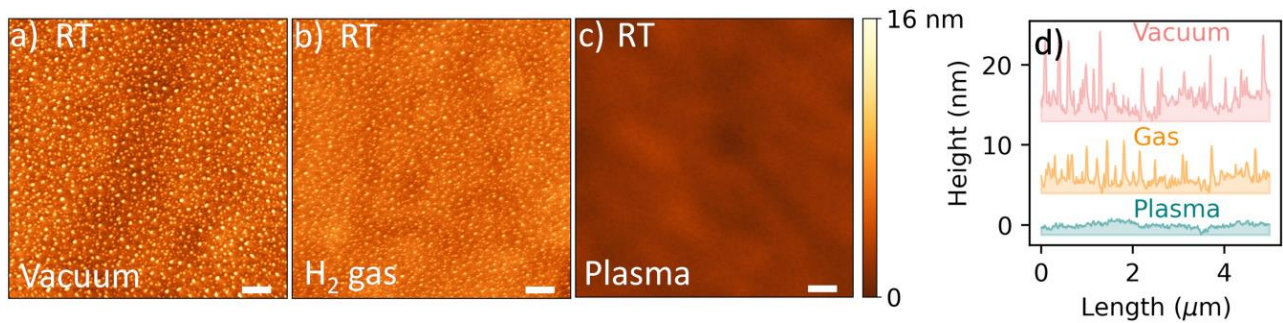

Figure 2: **Ion beam irradiation of GaP at room temperature in vacuum, H$_2$ gas, and hydrogen plasma environments. (a-c)** AFM maps of three regions of GaP that had been irradiated by ions in (a) vacuum, (b) H$_2$ gas and (c) hydrogen plasma environments. **(d)** AFM line scans showing the roughness of the surfaces imaged in (a-c). Ga droplet formation is not affected by H$_2$ gas, but it is suppressed by the hydrogen plasma. The ion beam energy and current were 12 keV and 2.3 nA.

The droplet formation seen in Figure 1 and 2(a) is a well-established consequence of ion beam restructuring of Ga-containing semiconductors.[15, 9] We therefore use Xe$^+$ irradiation of GaP as a well-established model system to investigate the role of hydrogen in suppressing material restructuring by ions.

## Ion beam irradiation in the presence of hydrogen gas and plasma

Figure 2 (b) and (c) show AFM maps of two regions of GaP that had been irradiated by 12 keV Xe$^+$ ions in H$_2$ gas and hydrogen plasma environments, respectively. The maps and the corresponding AFM line profiles in Figure 2(d) show that H$_2$ gas has a negligible effect on droplet formation, whilst a hydrogen plasma can eliminate droplet formation entirely. We attribute the ineffectiveness of H$_2$ gas at suppressing surface restructuring to the high stability of H$_2$ molecules. In contrast, the remote RF plasma[21] delivers highly reactive, thermalized hydrogen radicals to the surface, and the droplet suppression seen in Figure 2(c,d) proceeds through a chemical pathway.

We emphasize that both the H$_2$ gas and the plasma are delivered to the sample by the same microinjector (see Figure 1(a)). A remote RF generator is turned off or on during gas or plasma delivery, respectively, and the gas pressure in the specimen chamber is the same in both cases. The primary difference between the two scenarios is that some hydrogen gas molecules are decomposed and excited by the RF generator, producing H$_2$* and H* radicals (measurements performed versus sample bias indicate that hydrogen ions are not delivered to the sample surface due to ion neutralization inside the capillary used to deliver the plasma to the sample). The suppression of Ga droplet formation by the plasma observed in Figure 2 therefore cannot be explained by a physical effect such as a reduction in the ion beam current density or ion energy caused by scattering of the ions by gas molecules. It is, instead, a chemical effect driven by hydrogen radicals.

To elucidate the chemical pathway, we start by eliminating the most common effect discussed in the FIB literature – namely, the formation of volatile gas molecules during ion irradiation of a solid [8]. In the case of GaP irradiation in the presence of hydrogen (gas or plasma), the volatile species would need top be $GaH_3$ and/or $Ga_2H_6$. The formation of such volatile molecules can, in principle, lead to an increase in the removal rate of Ga, and thus cause the suppression of Ga droplet formation seen in Figure 2(c). However, $GaH_3$ and $Ga_2H_6$ are both extremely unstable. Their formation is energetically unfavorable and they decompose rapidly to form Ga and $H_2$.[22, 23, 24, 25] Volatilization of Ga is therefore highly unlikely, and so is desorption of $GaH_3$ and $Ga_2H_6$ as the reason for the absence of Ga droplets seen in Figure 2(c). We therefore disregard volatilization as a possibility and instead consider the effects of hydrogen on defects generated by ion irradiation, and resulting consequences for sputtering of GaP and Ga by ions.

Physical sputtering is caused primarily by momentum transfer from the keV ions to atoms that make up a crystal, and the sputtering yield is influenced by the masses and binding energies of the atoms. The binding energies are modified (typically reduced) at point defects such as vacancies created by ion impacts,[26, 8] and modified further by the addition of hydrogen that forms chemical bonds at the defects.[27, 28] In semiconductors, hydrogen is known to stabilize defects such as vacancies,[29, 30] and a hydrogen plasma has been shown previously to reduce the sputter rate of Ge during irradiation by 12 keV $Xe^+$ ions (at 400 °C).[21] In metals, the opposite effect is expected since hydrogen typically destabilizes vacancy defects.[31, 32] Here, we cannot measure unambiguously the effect of hydrogen on the sputter rate of GaP at room temperature due to the accumulation of metallic Ga at the surface which itself modifies the sputter rate. We can, however, show the effect of the plasma on sputtering of Ga droplets, as is done in Figure 3(a). Image (i) shows Ga droplets that had been generated by Xe+ irradiation of GaP in vacuum, and images (ii) - (iv) show that the droplets are removed upon introduction of the plasma (during continued ion irradiation). The plasma increases the removal rate of Ga, consistent with the argument that hydrogen destabilizes defects in metallic Ga and thus increases the sputter rate of the droplets (we note that the plasma does not erode the Ga droplets in the absence of ion irradiation).

To clarify the role of hydrogen further, we repeated the experiment in Figure 3(a) at an elevated sample temperature of 250 °C (Figure 3(b)). The elevated temperature has two main consequences. The Ga melts (whilst remaining stable in vacuum since the melting and boiling points are 30 °C and 2,400 °C, respectively), and the Ga droplets become mobile and diffuse along the surface. The droplet mobility is illustrated by Supplementary Video S2 in the Supplementary Information. It is also evident in Figure 3(b), image (i), due to the formation of trails by diffusing Ga droplets. The droplet mobility requires both an elevated temperature, and FIB irradiation (i.e., pausing the ion irradiation pauses the droplet motion).

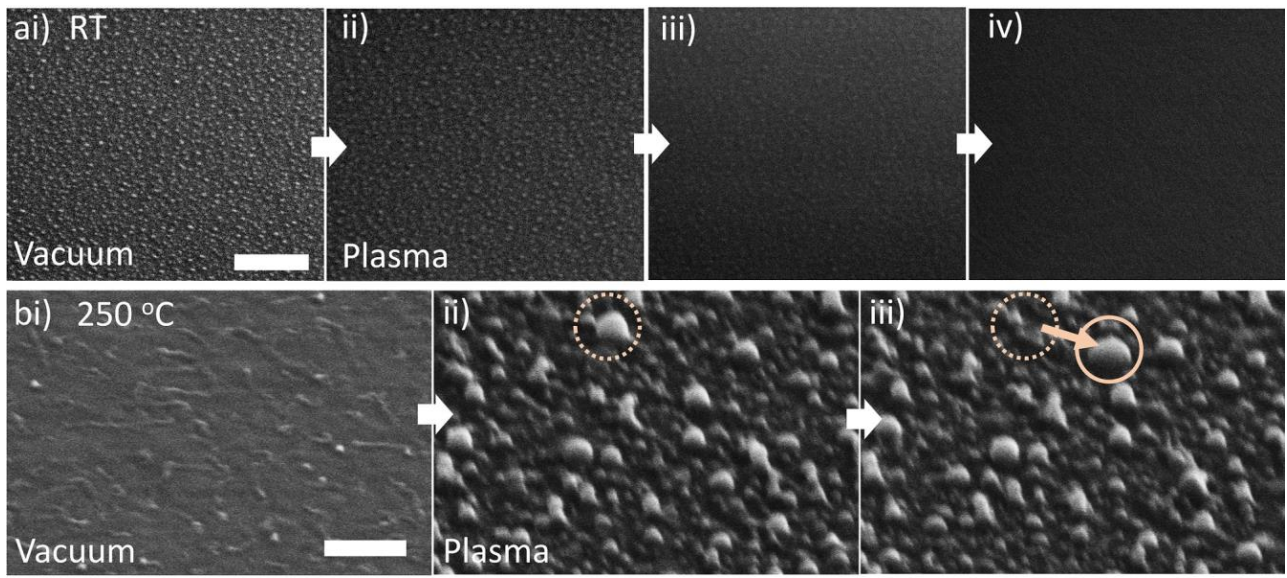

Figure 3: **Hybrid experiments showing the effects of a hydrogen plasma on excess Ga on the surface of GaP during ion beam irradiation. (a)** Time-resolved SEM image series showing Ga droplets generated by FIB irradiation in vacuum (i), and droplet erosion caused by the introduction of a H$_2$ plasma during the FIB irradiation (ii–iv). The experiment was performed at room temperature. The scale bar represents 500 nm. **(b)** Analogous experiment performed at 250 °C, where the Ga droplets are mobile in both vacuum and hydrogen plasma

environments. The scale bar represents 400 nm. A mobile Ga droplet is highlighted in frames (ii) and (iii). The Xe+ ion beam energy and current were 12 keV and 2.5 nA.

Now, in contrast to the behavior observed at room temperature (Figure 3(a)), introduction of the plasma at 250 °C (Figure 3(b)) does not increase the sputter rate of the Ga droplets. It instead causes an increase in the size of the Ga droplets which continue to diffuse along the surface, as is seen in images (ii) and (iii) of Figure 3(b), and in Supplementary Video S2. This is consistent with our proposed role of hydrogen at room temperature – that is, hydrogen accelerates the sputter rate of Ga at room temperature (Figure 3(a)) due to modification of atom binding energies at crystallographic defects which do not exist in liquid Ga at 250 °C. The increase in droplet size seen in 3(b) is likely caused by an increase in the mobility of excess Ga at the surface by hydrogen.

To summarize thus far, the key finding is that a hydrogen plasma suppresses the restructuring of GaP and prevents the formation of Ga droplets during ion beam irradiation at room temperature (Figure 2). The proposed mechanism is incorporation of hydrogen at crystallographic defects, which increases defect binding energies and thus inhibits defect migration and clustering within the ion-solid interaction volume. This mechanism is consistent with the effects of the plasma on pre-existing Ga droplets (Figure 3). It is also consistent with prior work on ion beam milling of GaP, GaAs and Ge, performed in the presence of a hydrogen plasma, at high temperatures (≥ 350 °C) [21]. In that work, the investigated materials were maintained above their recrystallisation temperatures. This is a special condition at which defects generated by the ions are annihilated by dynamic annealing, the materials remain crystalline during ion bombardment, and the plasma prevents surface roughening by immobilizing vacancies at the surfaces of semiconductors [21].

## Ion beam irradiation in the presence of $H_2O$

Finally, we repeated our plasma-assisted FIB experiments using $H_2O$ vapor in place of hydrogen. This control experiment is important because $H_2O$ is the primary contaminant present in high vacuum FIB sample chambers, and in gas injection systems such as the one we used to deliver hydrogen gas/plasma to the sample. The role of $H_2O$ must therefore be understood, delineated and accounted for in order to confirm our claims pertaining to hydrogen, and to eliminate the potential roles of alternative chemical pathways based on $H_2O$.

Figures 4(a) and (b) are AFM images of two regions of GaP that had been irradiated by 12 keV $Xe^+$ ions at room temperature in vacuum and in the presence of $H_2O$ vapor, respectively. Ga droplets are generated in vacuum, as before, but they are not generated in a $H_2O$ environment. The reason for the latter is, however, different from that of the hydrogen plasma – $H_2O$ gives rise to the formation of a surface oxide layer, as is illustrated by two experiments shown in Figure 4(c) and (d).

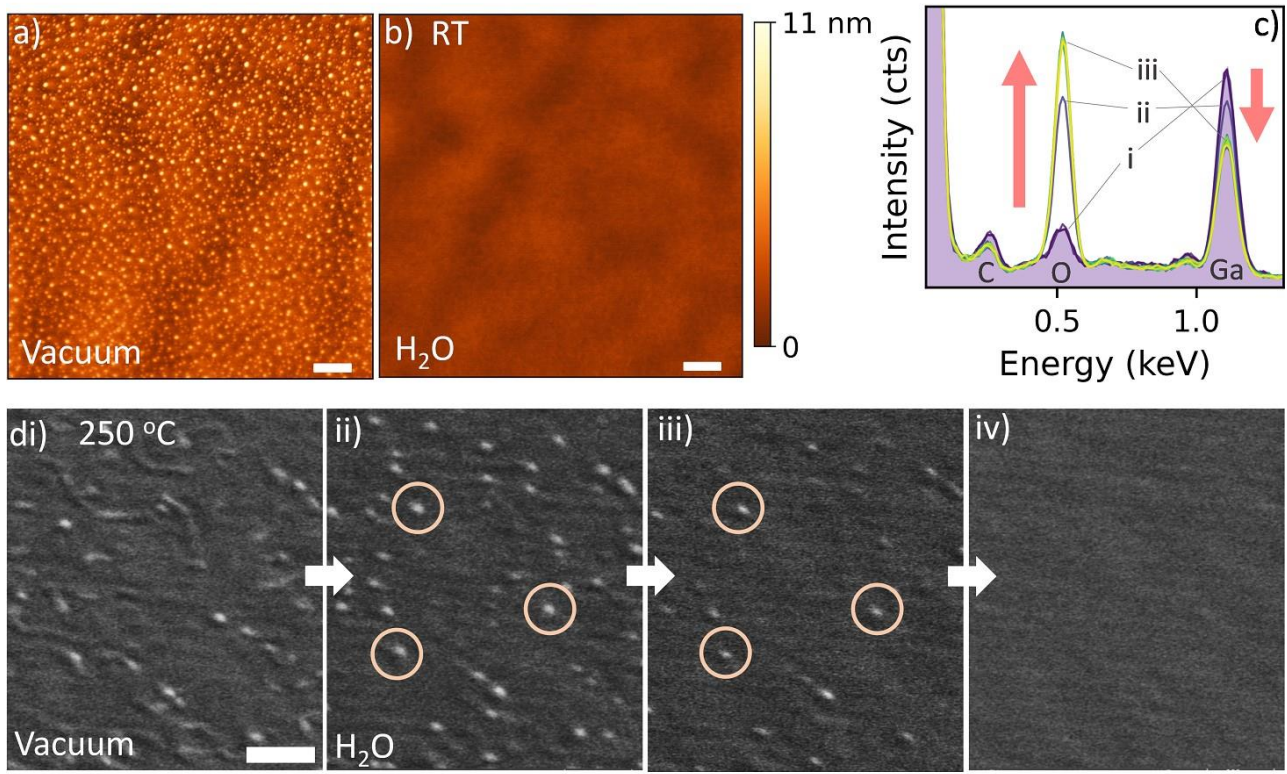

Figure 4: **Effects of H$_2$O vapor on excess Ga accumulation and ablation on the surface of GaP during ion beam irradiation. (a,b)** AFM maps of two regions of GaP that had been irradiated by ions in vacuum and in a H$_2$O vapor environment. **(c)** Timeresolved room temperature EDS spectra of a GaP surface in vacuum before ion irradiation (i), during H$_2$O exposure before ion irradiation (ii), and during H$_2$O exposure after ion irradiation (iii). The spectra show that H$_2$O gives rise to oxide formation. The Xe$^+$ ion beam energy and current were 12 keV and 0.83 nA. **(d)** Time-resolved SEM image series showing mobile Ga droplets generated by FIB irradiation in vacuum (i), and droplet immobilization and erosion caused by the introduction of H$_2$O vapor during the FIB irradiation (ii–ii). The experiment was performed at 250 °C. The scale bar represents 300 nm. The Xe$^+$ ion beam energy and current were 12 keV and 0.83 nA.

Figure 4(c) shows EDS spectra of a GaP surface (with a native oxide layer) measured in: (i) vacuum, before ion irradiation, (ii) H$_2$O vapor before ion irradiation, and (iii) H$_2$O vapor after ion irradiation. In (ii), the intensity of the OK$_\alpha$ x-ray peak increased (during EDS analysis) due to electron-beam-induced oxidation of the surface. In (iii), the OK$_\alpha$ intensity increased further due to oxidation caused by ion irradiation in the presence of H$_2$O vapor. The dramatic increase in the OK$_\alpha$ peak intensity seen in Figure 4(c) is absent when EDS analysis and ion irradiation are performed in vacuum or in a hydrogen plasma environment (see Figure S1 of the Supplementary Information). The oxide formation is caused by oxygen liberated through electron- and ion-induced dissociation of surface-adsorbed H$_2$O molecules.[33, 34, 35, 36]

Oxide formation at the GaP surface is important in the context of this work for two reasons. First, gallium oxide has a lower surface binding energy[37, 38, 39] (i.e., a higher sputtering yield) than Ga. Second, H$_2$O adsorbates are present on all surfaces in high vacuum ($\sim 10^{-4}$ Pa) FIB sample chambers. It may therefore be speculated that H$_2$O plays a role in our plasma experiments because H$_2$O contaminants may be present in the H$_2$ gas, and/or the hydrogen plasma may decompose H$_2$O adsorbates present at the sample surface and thus give rise to oxide formation. The EDS spectra in Figure 4(c) and Figure S1 suggest that this is not the case, since FIB irradiation in the presence of a hydrogen plasma causes a slight reduction rather than an increase in the intensity of the OK$_\alpha$ x-ray peak (Figure S1(b,c)). These EDS results indicate that $H_2O$ is not responsible for the chemical effects (in Figure 2 and 3) that we attribute to hydrogen.

Nonetheless, to confirm this further, we performed the additional test shown in Figure 4(d). In this experiment: (i) mobile, diffusing Ga droplets were generated on the surface of GaP by FIB irradiation in vacuum at 250 °C, and (ii)-(iv) $H_2O$ vapor was introduced as the FIB irradiation was continued. The introduction of $H_2O$ causes rapid immobilization of the diffusing droplets (images (ii) and (iii)), followed by erosion of the droplets by the ions (images (iii) and (iv)). This behavior is illustrated further by Supplementary Video S3 of the Supplementary Information. The droplet immobilization is consistent with gallium oxide formation since Ga is a liquid whilst gallium oxide is a solid at 250 °C. This is distinctly different from the case of the hydrogen plasma (shown in Figure 3(b) and Movie S2). The plasma does not cause droplet immobilization, nor does it cause droplet erosion at 250 °C. This result therefore confirms that the hydrogen plasma does not cause oxidation due to the presence of $H_2O$ contaminants in the $H_2$ gas, or in the gas delivery system.

Finally, we note that the droplet immobilization effect shown in Figure 4(d) provides a very sensitive, rapid means to detect gallium oxide formation. It is more sensitive and faster than EDS analysis because the electron beam dose needed to generate EDS spectra is much greater than that needed to observe droplets in SEM images. In fact, we use the SEM imaging method to test the purity of the $H_2$ used in our experiments. We found that active purification using a liquid nitrogen cold trap is needed to ensure that $H_2O$ contaminants do not dominate chemical effects at the GaP surface, even when using a high purity (≥ 99.99%) source of $H_2$, as is detailed in the Methods.

# Conclusion

We presented a minimally-intrusive chemical method for surface stabilisation during ion beam processing of materials, demonstrated using the III-V compound semiconductor GaP. The method entails the injection of thermalized hydrogen radicals to the surface during ion irradiation. It prevents changes in surface stoichiometry and topography caused by ion beam damage which has, to date, limited the applicability of ion beam processing techniques.

# Methods

**Materials and irradiation parameters:** 10×10 mm substrates were cleaved from a ⟨100⟩ orientated GaP wafer (MTI) and sonicated in acetone and isopropanol for 15 min each, and gently purged with flowing $N_2$. Individual samples for each dataset were loaded into a Thermo Fisher Scientific HELIOS G4 dual (ion-electron) beam microscope and pumped to a base pressure of 9 x $10^{-7}$ mbar before each experiment. All FIB irradiations were performed using a focused, 12 keV $Xe^+$ beam, and the beam currents specified in figure captions. Each irradiation was performed by scanning the beam over an area of 30 × 30 $\mu$m, with a dwell time and pixel overlap of 1 $\mu$sec and 50%, respectively. Elevated temperature (250 °C) FIB irradiations were performed using a custom-built boron nitride restive heating stage.[21]

**Analysis of surface Topography and elemental composition:** Secondary electron imaging was performed using the HELIOS G4 SEM, either in real time during FIB irradiation, or in an interlaced fashion whereby the FIB irradiation was paused periodically, whilst SEM images were collected. The electron beam was coincident with the FIB at the sample surface, the sample was at normal incidence with the FIB, and tilted 52 ° with respect to the electron beam, as is shown in Figure 1(a). *In-situ* x-ray analysis was performed using an Ultim Max x-ray detector (Oxford Instruments), and an electron beam energy of 3 keV (Figure 1) or 2 keV (3.2 nA) (Figure 4). *Ex-situ* AFM characterisation was performed using a Park XE7 AFM.

**H₂ gas/plasma delivery:** Hydrogen gas and plasma species were delivered to the sample using a home-built microinjector.[21] A gas/plasma injection capillary was aligned to the field of view of the FIB/SEM, at a distance of approximately 200 $\mu$m above the sample surface. H₂ gas from an ultra-pure lecture bottle was further dried using a liquid nitrogen cold-trap and a water-specific molecular sieve to ensure delivery of dry H₂ to the microinjector. An RF plasma was generated using a power of 64 W, and operated at a FIB sample chamber pressure of ~ $6 \times 10^{-5}$ mbar. H₂ gas experiments were performed at the same chamber pressure.

**H₂O vapour delivery:** H₂O vapor was delivered to the sample in a similar fashion using a commercial microinjector (Thermo Fisher Multichem GIS). Experiments were performed with a chamber pressure of $4 \times 10^{-5}$ mbar.

Additional information on conditions used to acquire the data in Figures 1-4 is provided in Section S4 of the Supplementary Information.

# Acknowledgement

We acknowledge financial support from the Australian Research Council (CE200100010).

# Supporting Information Available

**S1**: Surface elemental analysis during ion milling under different environments.

**S2**: Ga droplet diffusion behavior in the presence of H₂O and H₂.

**S3: Supplementary Videos (captions)**.

**S4: Extended Methods**.

**Supplementary Video S1**: Ga droplet growth during room temperature, high vacuum milling.

**Supplementary Video S2**: Surface topography at elevated temperatures with the introduction of $H_2$ gas followed by $H_2$ plasma.

**Supplementary Video S3**: Surface topography at elevated temperatures before and after the introduction of $H_2O$ vapour.

# References

[1] Li, P., Chen, S., Dai, H., Yang, Z., Chen, Z., Wang, Y., Chen, Y., Peng, W., Shan, W., and Duan, H., Recent advances in focused ion beam nanofabrication for nanostructures and devices: Fundamentals and applications, *Nanoscale* 13(3) (**2021**), pp. 1529–1565.

[2] Fröch, J. E., Bahm, A., Kianinia, M., Mu, Z., Bhatia, V., Kim, S., Cairney, J. M., Gao, W., Bradac, C., Aharonovich, I., and Toth, M., Versatile direct-writing of dopants in a solid state host through recoil implantation, *Nature Communications* 11(1) (**2020**), p. 5039.

[3] Berger, C., Premaraj, N., Ravelli, R. B., Knoops, K., López-Iglesias, C., and Peters, P. J., Cryo-electron tomography on focused ion beam lamellae transforms structural cell biology, *Nature Methods* 20(4) (**2023**), pp. 499–511.

[4] Zhang, Z., Wang, W., Dong, Z., Yang, X., Liang, F., Chen, X., Wang, C., Luo, C., Zhang, J., Wu, X., Sun, L., and Chu, J., The trends of in situ focused ion beam technology: toward preparing transmission electron microscopy lamella and devices at the atomic scale, *Advanced Electronic Materials* 8(9) (**2022**), p. 2101401.

[5] Höflich, K., Hobler, G., Allen, F. I., Wirtz, T., Rius, G., Krasheninnikov, A. V., Schmidt, M., Utke, I., Klingner, N., Osenberg, M., Córdoba, R., Djurabekova, F., Manke, I., Moll, P., Manoccio, M., De Teresa, J. M., Bischoff, L., Michler, J., De Castro, O., Delobbe, A., Dunne, P., Dobrovolskiy, O. V., Frese, N., Gölzhäuser, A.,

Mazarov, P., Koelle, D., Mo¨ller, W., P´erez-Murano, F., Philipp, P., Vollnhals, F., and Hlawacek, G., Roadmap for focused ion beam technologies, *Applied Physics Reviews* 10(4) (**2023**), pp. 1931–9401.

[6]  Facsko, S., Dekorsy, T., Koerdt, C., Trappe, C., Kurz, H., Vogt, A., and Hartnagel, H. L., Formation of ordered nanoscale semiconductor dots by ion sputtering, *Science* 285(5433) (**1999**), pp. 1551–1553.

[7]  Chan, W. L. and Chason, E., Making waves: kinetic processes controlling surface evolution during low energy ion sputtering, *Journal of Applied Physics* 101(12) (**2007**), p. 121301.

[8]  Behrisch, R. and Eckstein, W., *Sputtering by particle bombardment: experiments and computer calculations from threshold to MeV energies*, vol. 110, Springer Science & Business Media, **2007**.

[9]  Kang, M. and Goldman, R., Ion irradiation of III–V semiconductor surfaces: From selfassembled nanostructures to plasmonic crystals, *Applied Physics Reviews* 6(4) (**2019**), p. 041307.

[10] Va´zquez, L., Redondo-Cubero, A., Lorenz, K., Palomares, F., and Cuerno, R., Surface nanopatterning by ion beam irradiation: compositional effects, *Journal of Physics: Condensed Matter* 34(33) (**2022**), p. 333002.

[11] Goldstein, J. I., Newbury, D. E., Michael, J. R., Ritchie, N. W., Scott, J. H. J., and Joy, D. C., *Scanning electron microscopy and X-ray microanalysis*, springer, **2017**.

[12] Ziegler, J. F., Ziegler, M. D., and Biersack, J. P., SRIM–The stopping and range of ions in matter (2010), *Nuclear Instruments and Methods in Physics Research Section B: Beam Interactions with Materials and Atoms* 268(11) (**2010**), pp. 1818–1823.

[13] MoberlyChan, W. J., Adams, D. P., Aziz, M. J., Hobler, G., and Schenkel, T., Fundamentals of focused ion beam nanostructural processing: Below, at, and above the surface, *MRS Bulletin* 32(5) (**2007**), pp. 424–432.

[14] Lugstein, A., Basnar, B., and Bertagnolli, E., Study of focused ion beam response of GaAs in the nanoscale regime, *Journal of Vacuum Science & Technology B: Microelectronics and Nanometer Structures Processing, Measurement, and Phenomena* 20(6) (**2002**), pp. 2238–2242.

[15] Grossklaus, K. and Millunchick, J., Mechanisms of nanodot formation under focused ion beam irradiation in compound semiconductors, *Journal of Applied Physics* 109(1) (**2011**), p. 014319.

[16] Wu, J., Ye, W., Cardozo, B., Saltzman, D., Sun, K., Sun, H., Mansfield, J., and Goldman, R., Formation and coarsening of Ga droplets on focused-ion-beam irradiated GaAs surfaces, *Applied Physics Letters* 95(15) (**2009**), p. 153107.

[17] Botman, A., Bahm, A., Randolph, S., Straw, M., and Toth, M., Spontaneous growth of gallium-filled microcapillaries on ion-bombarded GaN, *Physical Review Letters* 111(13) (**2013**), p. 135503.

[18] Wei, Q., Lian, J., Lu, W., and Wang, L., Highly ordered Ga nanodroplets on a GaAs surface formed by a focused ion beam, *Physical Review Letters* 100(7) (**2008**), p. 076103.

[19] Kang, M., Wu, J., Huang, S., Warren, M. V., Jiang, Y., Robb, E., and Goldman, R. S., Universal mechanism for ion-induced nanostructure formation on III-V compound semiconductor surfaces, *Applied Physics Letters* 101(8) (**2012**), p. 082101.

[20] Tang, S.-Y., Tabor, C., Kalantar-Zadeh, K., and Dickey, M. D., Gallium liquid metal: the devil's elixir, *Annual Review of Materials Research* 51 (**2021**), pp. 381–408.

[21] Scott, J. A., Bishop, J., and Toth, M., Suppression of Surface Roughening during Ion Bombardment of Semiconductors, *Chemistry of Materials* 34(19) (**2022**), pp. 8968–8974.

[22] Pulham, C. R., Downs, A. J., Goode, M. J., Rankin, D. W., and Robertson, H. E., Gallane: synthesis, physical and chemical properties, and structure of the gaseous molecule $Ga_2H_6$ as determined by electron diffraction, *Journal of the American Chemical Society* 113(14) (**1991**), pp. 5149–5162.

[23] Andrews, P. C., Gardiner, M. G., Raston, C. L., and Tolhurst, V.-A., Structural aspects of tertiary amine adducts of alane and gallane, *Inorganica Chimica Acta* 259(1-2) (**1997**), pp. 249–255.

[24] Conn, R., Doerner, R., Sze, F., Luckhardt, S., Liebscher, A., Seraydarian, R., and Whyte, D., Deuterium plasma interactions with liquid gallium, *Nuclear Fusion* 42(9) (**2002**), p. 1060.

[25] Downs, A. J., Greene, T. M., Johnsen, E., Pulham, C. R., Robertson, H. E., and Wann, D. A., The digallane molecule, $Ga_2H_6$: experimental update giving an improved structure and estimate of the enthalpy change for the reaction $Ga_2H_6$ (g) $\rightarrow$ $2GaH_3$ (g), *Dalton Transactions* 39(24) (**2010**), pp. 5637–5642.

[26] Sigmund, P., Theory of sputtering. I. Sputtering yield of amorphous and polycrystalline targets, *Physical Review* 184(2) (**1969**), p. 383.

[27] Horn, A., Schenk, A., Biener, J., Winter, B., Lutterloh, C., Wittmann, M., and Küppers, J., H atom impact induced chemical erosion reaction at C: H film surfaces, *Chemical Physics Letters* 231(2-3) (**1994**), pp. 193–198.

[28] Jacob, W. and Roth, J., Chemical sputtering, *Sputtering by Particle Bombardment: Experiments and Computer Calculations from Threshold to MeV Energies*, Springer, **2007**, pp. 329–400.

[29] Jones, R., Coomer, B., Goss, J. P., Hourahine, B., and Resende, A., The interaction of hydrogen with deep level defects in silicon, *Solid State Phenomena* 71 (**1999**), pp. 173–248.

[30] Pearton, S. J., Corbett, J. W., and Stavola, M., *Hydrogen in crystalline semiconductors*, vol. 16, Springer Science & Business Media, **2013**.

[31] Du, J.-P., Geng, W., Arakawa, K., Li, J., and Ogata, S., Hydrogen-enhanced vacancy diffusion in metals, *The Journal of Physical Chemistry Letters* 11(17) (**2020**), pp. 7015–7020.

[32] Huang, L., Chen, D., Xie, D., Li, S., Zhang, Y., Zhu, T., Raabe, D., Ma, E., Li, J., and Shan, Z., Quantitative tests revealing hydrogen-enhanced dislocation motion in $\alpha$-iron, *Nature Materials* (**2023**), pp. 1–7.

[33] Ebinger, H., Yates, J., et al., Electron-impact-induced oxidation of Al (111) in water vapor: Relation to the Cabrera-Mott mechanism, *Physical Review B* 57(3) (**1998**), p. 1976.

[34] Geier, B., Gspan, C., Winkler, R., Schmied, R., Fowlkes, J. D., Fitzek, H., Rauch, S., Rattenberger, J., Rack, P. D., and Plank, H., Rapid and highly compact purification for focused electron beam induced deposits: a low temperature approach using electron stimulated $H_2O$ reactions, *The Journal of Physical Chemistry C* 118(25) (**2014**), pp. 14009–14016.

[35] Walia, S., Balendhran, S., Ahmed, T., Singh, M., El-Badawi, C., Brennan, M. D., Weerathunge, P., Karim, M. N., Rahman, F., Rassell, A., Duckworth, J., Ramanathan, R.,

Collis, G. E., Lobo, C. J., Toth, M., Kotsakidis, J. C., Weber, B., Fuhrer, M., Dominguez-Vera, J. M., Spencer, M. J. S., Aharonovich, I., Sriram, S., Bhaskaran, M., and Bansal, V., Ambient protection of few-layer black phosphorus via sequestration of reactive oxygen species, *Advanced Materials* 29(27) (**2017**), p. 1700152.

[36] Rummeli, M. H., Ta, H. Q., Mendes, R. G., Gonzalez-Martinez, I. G., Zhao, L., Gao, J., Fu, L., Gemming, T., Bachmatiuk, A., and Liu, Z., New frontiers in electron beam–driven chemistry in and around graphene, *Advanced Materials* 31(9) (**2019**), p. 1800715.

[37] Egry, I., Ricci, E., Novakovic, R., and Ozawa, S., Surface tension of liquid metals and alloys—Recent developments, *Advances in Colloid and Interface Science* 159(2) (**2010**), pp. 198–212.

[38] Xu, Q., Oudalov, N., Guo, Q., Jaeger, H. M., and Brown, E., Effect of oxidation on the mechanical properties of liquid gallium and eutectic gallium-indium, *Physics of Fluids* 24(6) (**2012**), p. 063101.

[39] Handschuh-Wang, S., Gan, T., Wang, T., Stadler, F. J., and Zhou, X., Surface tension of the oxide skin of gallium-based liquid metals, *Langmuir* 37(30) (**2021**), pp. 9017–9025.